\documentclass[sigconf, nonacm]{acmart}
\usepackage{multirow}
\usepackage{pgfplots}
\usepackage{pifont}
\pgfplotsset{compat=1.18}
\usepackage{caption}
\usepackage[ruled,vlined,linesnumbered,noend]{algorithm2e}

\newcommand\vldbdoi{XX.XX/XXX.XX}
\newcommand\vldbpages{XXX-XXX}
\newcommand\vldbvolume{14}
\newcommand\vldbissue{1}
\newcommand\vldbyear{2020}
\newcommand\vldbauthors{\authors}
\newcommand\vldbtitle{\shorttitle} 
\newcommand\vldbavailabilityurl{URL_TO_YOUR_ARTIFACTS}
\newcommand\vldbpagestyle{plain} 

\begin{document}
\title{Rel-HNN: Split Parallel Hypergraph Neural Network for Learning on Relational Databases}

\author{Md. Tanvir Alam}
\affiliation{%
  \institution{University of Dhaka}
  \streetaddress{P.O. Box 1212}
  \postcode{43017-6221}
}
\email{tanvir15@du.ac.bd}

\author{Md. Ahasanul Alam}
\affiliation{%
  \institution{Brac University}
  \streetaddress{P.O. Box 1212}
  \postcode{43017-6221}
}
\email{ahasanul.alam@bracu.ac.bd}

\author{Md Mahmudur Rahman}
\affiliation{%
  \institution{University of Dhaka}
  \streetaddress{P.O. Box 1212}
  \postcode{43017-6221}
}
\email{mahmudur@cse.du.ac.bd}

\author{Md. Mosaddek Khan}
\affiliation{%
  \institution{University of Dhaka}
  \streetaddress{P.O. Box 1212}
  \postcode{43017-6221}
}
\email{mosaddek@du.ac.bd}

\settopmatter{authorsperrow=4}

\begin{abstract}

Relational databases (RDBs) are ubiquitous in enterprise and real-world applications. Flattening the database poses challenges for deep learning models that rely on fixed-size input representations to capture relational semantics from the structured nature of relational data. Graph neural networks (GNNs) have been proposed to address this, but they often oversimplify relational structures by modeling all the tuples as monolithic nodes and ignoring intra-tuple associations. In this work, we propose a novel hypergraph-based framework, that we call rel-HNN, which models each unique attribute-value pair as a node and each tuple as a hyperedge, enabling the capture of fine-grained intra-tuple relationships. Our approach learns explicit multi-level representations across attribute-value, tuple, and table levels. To address the scalability challenges posed by large RDBs, we further introduce a split-parallel training algorithm that leverages multi-GPU execution for efficient hypergraph learning. Extensive experiments on real-world and benchmark datasets demonstrate that rel-HNN significantly outperforms existing methods in both classification and regression tasks. Moreover, our split-parallel training achieves substantial speedups—up to 3.18$\times$ for learning on relational data and up to 2.94$\times$ for hypergraph learning—compared to conventional single-GPU execution.

\end{abstract}

\maketitle

\pagestyle{\vldbpagestyle}
\begingroup\small\noindent\raggedright\textbf{PVLDB Reference Format:}\\
\vldbauthors. \vldbtitle. PVLDB, \vldbvolume(\vldbissue): \vldbpages, \vldbyear.\\
\href{https://doi.org/\vldbdoi}{doi:\vldbdoi}
\endgroup
\begingroup
\renewcommand\thefootnote{}\footnote{\noindent
This work is licensed under the Creative Commons BY-NC-ND 4.0 International License. Visit \url{https://creativecommons.org/licenses/by-nc-nd/4.0/} to view a copy of this license. For any use beyond those covered by this license, obtain permission by emailing \href{mailto:info@vldb.org}{info@vldb.org}. Copyright is held by the owner/author(s). Publication rights licensed to the VLDB Endowment. \\
\raggedright Proceedings of the VLDB Endowment, Vol. \vldbvolume, No. \vldbissue\ %
ISSN 2150-8097. \\
\href{https://doi.org/\vldbdoi}{doi:\vldbdoi} \\
}\addtocounter{footnote}{-1}\endgroup

\ifdefempty{\vldbavailabilityurl}{}{
\vspace{.3cm}
\begingroup\small\noindent\raggedright\textbf{PVLDB Artifact Availability:}\\
The source code, data, and/or other artifacts have been made available at \url{https://github.com/tfahim15/rel-HNN}.
\endgroup
}

\section{Introduction}

Relational databases (RDBs) are among the most widely used forms of data representation in enterprise environments, owing to their ability to efficiently handle structured data, support complex queries, and maintain data integrity. A relational database consists of multiple tables governed by a schema that defines the relationships among them. RDBs serve as the primary data format across a wide range of industries, including online advertising, recommender systems, healthcare, and fraud detection.

Despite their widespread use, the direct application of machine learning—particularly deep learning—to relational databases (RDBs) has received limited attention from the research community. Traditionally, applying machine learning to relational databases requires transforming the data into a flat, tabular format, since most supervised learning models rely on fixed-size input vectors. This transformation process, commonly referred to as \textit{flattening}, typically involves joining multiple related tables into a single denormalized table with predefined columns. Flattening usually demands extensive, rule-based feature engineering~\cite{ruleBased_Feature1, ruleBased_Feature2, ruleBased_Feature3}, and frequently results in the loss of valuable relational information embedded within the schema and data. Moreover, for large-scale RDBs, flattening introduces substantial computational overhead, becoming a major bottleneck in the overall machine learning pipeline. Consequently, there is an increasing demand for machine learning approaches capable of directly operating on relational data in its native form, without the need for manual feature engineering or flattening.

Effective application of neural network-based techniques to relational databases hinges on overcoming two fundamental challenges. First, the inherently structured yet complex nature of relational databases, characterized by multiple interconnected tables, demands models capable of capturing and leveraging these intricate relationships. Second, the considerable scale of relational databases—often containing millions or even billions of records across dozens to hundreds of tables—necessitates efficient training and inference procedures to ensure both practicality and scalability~\cite{paper2}. A recent approach that has gained traction to address these challenges involves applying Graph Neural Networks (GNNs) directly to relational databases~\cite{gnn1, gnn2, gnn3, gnn4, gnn5, paper1}. In this context, the primary task typically involves predicting values for a target column in a specific table, using available relational information from the entire database. To achieve this, databases are modeled as graphs, where tuples serve as vertices and relationships between tuples from different tables—defined by foreign keys—serve as edges. To better capture relational semantics, existing methods have proposed utilizing relation-type-dependent weights~\cite{gnn6}, specialized convolution operators~\cite{gnn7}, or generative architectures~\cite{gnn8} within relational graphs.

Despite their expressive power, existing GNN-based approaches have notable limitations when modeling relational databases. Primarily, they treat entire tuples as monolithic nodes, thus ignoring the granular, attribute-level structures and failing to capture fine-grained inter-attribute interactions. Moreover, relying solely on primary key–foreign key (PK–FK) relationships significantly limits their ability to represent meaningful relationships among tuples within the same table. Additionally, these methods often overlook symmetries among sub-graphs and require multiple rounds of message passing, resulting in inefficient training and inference processes, particularly at scale. Furthermore, existing methods are heavily schema-dependent, making them challenging to generalize or adapt to new datasets without substantial human effort to define or extract PK–FK constraints explicitly.

An emerging alternative is hypergraph-based modeling, which has recently shown promise for automated learning on relational databases \cite{paper3}. A hypergraph generalizes traditional graphs by allowing edges (termed hyperedges) to connect an arbitrary number of vertices rather than being restricted to pairs. A hypergraph can reveal complex structural patterns involving multiple vertices and provide insights into network dynamics—patterns that traditional graphs, limited to pairwise connections, fail to capture
\cite{surveyHypergraph}. This ability to represent higher-order interactions among vertices has gained significant attention across various real-world complex systems, including physical systems~\cite{hypergraph_app1}, microbial communities~\cite{hypergraph_app2}, brain functions~\cite{hypergraph_app3}, and social networks~\cite{hypergraph_app4}. The flexibility of hypergraphs in modeling multi-way interactions has motivated the development of powerful hypergraph neural network (HGNN) algorithms~\cite{feng2019hypergraph, chien2022you, yadati2019hypergcn}, facilitating the learning of intricate relational patterns.

For relational databases specifically, ATJ-Net \cite{paper3} leverages a hypergraph to train a heterogeneous GNN. It initially represents joinable attributes as vertices and tuples as   hyperedges. The hypergraph is then transformed into to a heterogenous bipartite graph where the tuples and joinable attributes constitute vertices. Then, it applies a message-passing GNN to the bipartite graph to predict labels associated with tuples in the target table. Although ATJ-Net includes joinable attributes alongside tuples, it has several critical limitations. First, it considers only categorical, joinable attributes—typically those defined explicitly through primary key–foreign key (PK–FK) relationships—thereby confining the model to predefined relational paths and potentially missing complex attribute associations within tuples. Second, transforming the hypergraph into a bipartite graph inevitably flattens high-order relationships into pairwise edges, causing the loss of valuable higher-order interactions that could otherwise be effectively captured by hypergraph neural networks. Based on the above discussion, our work makes the following key contributions:

\begin{itemize}
    \item We propose a novel hypergraph representation for relational data that preserves both intra-tuple and inter-tuple relationships. Unlike traditional methods relying exclusively on primary key–foreign key (PK–FK) constraints, our representation decomposes tuples into attribute–value pairs, creating nodes naturally connected via hyperedges. This approach effectively captures fine-grained attribute-level interactions and is inherently schema-agnostic, eliminating the need for manual feature engineering or explicit schema knowledge. To the best of our knowledge, this is the first hypergraph-based representation specifically developed for relational database learning.
    
    \item We introduce rel-HNN, a hypergraph neural network specifically tailored for relational databases. Leveraging our hypergraph structure, rel-HNN learns explicit embeddings at three granularity levels—attribute–value pairs, tuples, and entire tables—thus effectively capturing both localized and global relational patterns and enabling richer relational learning.
    
    \item Additionally, to overcome the challenges posed by large-scale relational databases, we propose a split-parallel hypergraph learning algorithm that leverages multi-GPU parallelism. Our method enables full-hypergraph training by partitioning both data and computation across GPUs while preserving global structural context. Unlike mini-batch GNN training, which introduces redundant data movement and overlooks neighborhood completeness, our approach ensures efficient and context-aware learning. To the best of our knowledge, this is the first work to introduce split-parallelism for scalable training of hypergraph neural networks.

    \item Finally, our extensive experimental results demonstrate that the proposed multi-level representation framework enables rel-HNN to significantly outperform the state-of-the-art methods on both classification and regression tasks across a diverse set of real-world relational datasets. Moreover, our split-parallel training framework further delivers substantial performance improvements, achieving up to 3.18$\times$ speedup on large-scale relational datasets and up to 2.94$\times$ on benchmark hypergraph datasets. These results highlight both the effectiveness and scalability of our approach.
\end{itemize}

The remainder of the paper is organized as follows. Section 2 introduces the preliminaries and formally defines the problem. Section 3 reviews relevant related work. Section 4 details the proposed methodology. Section 5 presents the experimental setup, results, and analysis. Finally, Section 6 concludes the paper and outlines potential directions for future work.

\section{Background}
\label{sec:background}

A relational database (RDB) is defined as a collection of tables, denoted by $RDB = \{T^{1}, T^{2}, \dots, T^{n}\}$, where each $T^{k} \in RDB$ represents a table. Each table captures information about a specific entity type, with rows (tuples) corresponding to entity instances and columns representing attributes or features. Table columns may contain diverse data types, including numerical values, categorical text, timestamps, geographic coordinates, and multimedia content. We denote the $i$-th row of table $T^{k}$ as $T^k_{i}$, and the $j$-th column as $T^K_{:j}$. Let $Attr^{T^{k}}$ denote the set of attributes (i.e., columns) of table $T^{k}$. An RDB is referred to as “relational” because values in a column $T^{k}_{:j}$ of one table may refer to rows in another table $T^{m} \in RDB$. Such columns are known as foreign keys and serve as the basis for modeling inter-table relationships.

\begin{figure}[h]
    \centering
    \includegraphics[width=0.9\linewidth]{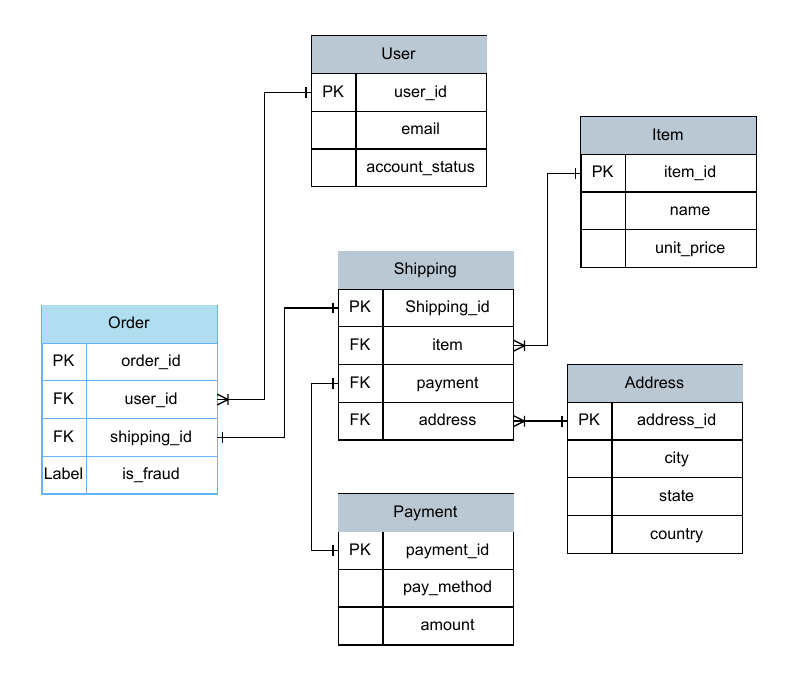}
    \caption{Schema Diagram of an example Relational Databse (RDB)}
    \label{fig:schema}
\end{figure}

Learning tasks on a relational database ($RDB$) are typically formulated as predicting the values of a specific column in a designated target table. Let $T^{tg} \in RDB$ be the target table. In the training data, the rows of table $T^{tg}$ in the training set are associated with a label. The goal is to predict the labels for the remaining rows of table $T^{tg}$  where the label values are unknown. Figure~\ref{fig:schema} illustrates a schema diagram of an example relational database. Here the table named $order$ is the target table, and the task is to learn and predict the label $is\_fraud$ for the rows.

\section{Related Works}
\label{sec:related_works}

In this section, we review research closely related to our work. Section~\ref{subsec:relw_graph} discusses graph-based learning approaches for relational data. Section~\ref{subsec:relw_dag} presents DAG-based models designed for single-pass learning. Section~\ref{subsec:relw_hypergraph} reviews recent efforts that apply hypergraph-based methods to relational databases, and Section~\ref{subsec:relw_dist_GNN} covers distributed multi-GPU training strategies for scalable graph and hypergraph learning.

\subsection{Learning on \textit{RDB} using Graphs}
\label{subsec:relw_graph}
The deep learning community has paid relatively little attention to learning directly from data in relational databases. The common approach in machine learning is to "flatten" relational data into a single table format, as most widely used supervised learning models require inputs in the form of fixed-size vectors. To reduce transfer overhead, frameworks such as MADlib integrate learning algorithms directly into the RDBMS as User Defined Aggregate Functions (UDAFs)\cite{learning_on_RDB_1,kara2021machine}. Research on \textit{learning over joins} aims to avoid full join output by exploiting relational constraints (e.g., functional dependencies) and factorized computation \cite{hai2023amalur, tziavelis2020optimal}. However, this flattening process often eliminates valuable relational information inherent in the data. Additionally, the feature engineering needed to perform this transformation is typically one of the most challenging and time-consuming tasks for machine learning practitioners.

\begin{table}[t]
    \centering
    
    \caption{Relational Database and Graph Terminology Mapping}
    \begin{tabular}{l l}
        \hline\textbf{Relational Database} & \textbf{Graph Representation} \\ \hline
         Row/Tuple &  Node \\
         Table & Node type \\
         Foreign key column & Edge type \\
         Non-foreign-key column & Node features \\
         Foreign key from $T^A_{u}$ to $T^B_{v}$ & Edge from node $u$ to node $v$\\ 
         Label of target tuple $t \in T^{tg}$ & Label of node for $t$ of type $tg$ \\
        \hline
        \\
    \end{tabular}
    \label{tab:table_1}
\end{table}

Recent approaches interpret the relational database as a Graph Neural Networks (GNN) that can accurately capture the relational structure in $RDBs$ \cite{cvitkovic2020supervised}. The relational database is transformed into a graph, representing the relational structure, to predict the target attribute. In the graph representation, each node corresponds to a tuple, and edges represent foreign key relationships between these tuples. To manage the complexity of the graph, connections are typically restricted to a specific depth or number of hops. Finally, a GNN on this graph is applied to predict the target attribute using the GNN. The relation between the $RDB$ and Graph representation is presented in Table \ref{tab:table_1}.

Representing a relational database (RDB) as a graph enables supervised learning tasks on RDBs to be framed as node classification problems~\cite{atwood_2016}. This representation supports both classification and regression tasks. The graph-based perspective also suggests that Graph Neural Network (GNN) techniques can be effectively applied to learning tasks involving relational databases. GNNs, particularly those designed for supervised learning on RDBs, commonly utilize the message passing framework~\cite{gilmer2017neural}. In this framework, each node $v$ in a graph is initially assigned a hidden representation $h_v^0$, which is iteratively updated over $R$ rounds of message passing. In each round $r$, node $v$ transmits a message $m_{vw}^r$ to each of its neighboring nodes $w$. These messages are generated by a learnable function that may incorporate edge features and depend on the current hidden states $h_v^r$ and $h_w^r$ of the source and target nodes, respectively. Each node then aggregates the messages received from its neighbors using another learnable function, resulting in an updated hidden state $h_v^{r+1}$. After $R$ rounds of message passing, a readout function aggregates the final hidden states $h_v^R$ across all nodes to produce a prediction for the entire graph.

\subsection{Using DAG for single pass learning}
\label{subsec:relw_dag}
Learning with GNNs typically requires multiple message passing among nodes, which leads to significantly high training and inference times. To address this limitation, a recent approach named SPARE introduces an alternative encoding technique that represents the $RDB$ as a directed acyclic graph (DAG)~\cite{paper2}. The DAG structure enables single-pass learning, allowing for faster training and inference compared to standard GNNs.

In SPARE, relational data is organized as a Directed Acyclic Graph (DAG), denoted by $DAG_t$, instead of using the undirected graphs typically employed by Graph Neural Networks (GNNs). This DAG is constructed by performing a breadth-first search (BFS) on the undirected graph $G_t$ rooted at a target tuple $t \in T^{tg}$. Edges are directed from nodes at greater depth to nodes at lesser depth. In cases where nodes share the same depth, a global ordering is used to determine edge directionality. This construction guarantees an acyclic structure with the target tuple positioned at the root, thereby defining a clear direction for message propagation. Additionally, SPARE introduces relational DAG pruning to reduce redundancy. Repeating sub-DAGs that appear across different target tuples are replaced with shared embeddings. This strategy compresses the graph and improves training efficiency by eliminating redundant computation.

SPARE’s learning procedure involves a single-pass message propagation over each constructed DAG. Each node (tuple) in $DAG_t$ is first embedded using a table-specific MLP to produce its initial hidden representation. The model then performs a bottom-up traversal of the DAG (following topological order), where each node aggregates messages from its children using a learnable aggregation function. The final hidden state of the root (target tuple) is passed through an output MLP to generate the prediction. This single-pass design allows SPARE to achieve significantly faster training and inference. However, both GNN- and DAG-based methods model relational data as standard graphs, often overlooking structural regularities defined by the database schema. This can lead to redundant computations and reduced learning efficiency. In contrast, a more recent approach, ATJ-Net, introduces a hypergraph-based representation that enables more adaptable and automated learning across relational databases~\cite{paper3}.

\subsection{Hypergraph-based learning}
\label{subsec:relw_hypergraph}
A hypergraph is a generalization of a traditional graph in which edges, called hyperedges, can connect any number of vertices, rather than being restricted to just two. A hypergraph can be represented as $H = (V, E, X)$, where $V = \{v_{1}, v_{2}, \dots, v_{|V|}\}$ is the set of nodes or vertices, $E= \{e_{1}, e_{2}, \dots , e_{|E|}\}$  denotes the set of hyperedges, and $X \in \mathbb{R}^{|V|\times d}$ is the feature matrix. Each hyperedge $e_i\subseteq V$ includes the set of nodes it connects. Following the success of graph neural networks, hypergraph neural network models have received growing attention. HGNN~\cite{feng2019hypergraph} and HyperGCN~\cite{yadati2019hypergcn} extend graph convolutional networks to the hypergraph setting, while AllSet~\cite{chien2022you} introduces a two-stage message-passing framework. In this approach, hyperedge representations are first computed by aggregating the embeddings of their constituent nodes from the previous layer; then, the embeddings of the hyperedges connected to a node are aggregated to update that node’s representation.

ATJ-Net\cite{paper3} models the relational database $RDB$ as a hypergraph, each tuple $T^k_t$ of table $T^k \in RDB$ is considered as a hyperedge, that is $E = \{T^k_i\}_{k,i}$ where $T^k_i$ denotes the $i-th$ tuple of table $T^k \in RDB$. Joinable attributes serve as the vertices, and other attributes are treated as features of the hyperedges. Every table—except the main table—must include at least one joinable attribute; otherwise, it cannot be linked to the target label. If a hyperedge contains a vertex, a connection is established between them. Formally, let $Attr^{T^k}(j)$ be the value set of $j-th$ joinable attribute of table $T^k$ and $V = \{Attr^{T^k}(j)\}_{k,j}$. Moreover, the hypergraph can be viewed as a bipartite graph, with tuples and joinable attributes represented as two distinct sets of vertices, and edges indicate their inclusion relationships.

ATJ-Net employs the message passing neural network (MPNN) framework~\cite{gilmer2017neural} to construct a GNN over the heterogeneous hypergraph. In each GNN layer, hyperedge features are first aggregated to the vertices, which are then updated and propagated back to the hyperedges. Although ATJ-Net uses a hypergraph-based formulation, it ultimately trains a GNN by stacking standard GNN layers, which limits its capacity to fully exploit the rich relational structures across tables in $RDB$. Additionally, it considers only joinable attributes and neglects the representation learning of other attributes, potentially reducing its effectiveness in modeling complex intra-tuple associations.

\subsection{Distributed multi-GPU GNN training}
\label{subsec:relw_dist_GNN}
The challenge of scaling to large graphs has led to extensive research on multi-GPU GNN training systems. Training strategies for utilizing multiple GPUs have been devised for both mini-batch and full-graph training \cite{10.1145/3514221.3526134,10.1145/3589288}. In mini-batch multi-GPU training, nodes are first sampled into micro-batches, one for each GPU. Then, the local $k$-hop neighborhoods of the target nodes in each micro-batch are sampled and loaded. Each GPU subsequently trains the model on its assigned micro-batch in a data-parallel fashion~\cite{wang2019deep,gandhi2021p3}. However, data-parallel training often results in redundant data loading and computation due to overlapping $k$-hop neighborhoods across micro-batches. To mitigate this issue, GSplit~\cite{Polisetty2025GSplit} partitions the mini-batch into non-overlapping splits, assigning each split to a specific GPU. Each GPU processes only the vertices in its assigned split and exchanges intermediate results at each GNN layer. Nonetheless, mini-batch GNN training relies on neighborhood sampling, which may exclude important neighbors, leading to information loss and suboptimal message propagation. 

Full-graph multi-GPU training~\cite{10.1145/3514221.3526134,10.1145/3589288}, in contrast, maintains the complete graph structure during training. G3~\cite{10.1145/3589288} splits the full graph into non-overlapping partitions, with each GPU responsible for one partition. After each layer, GPUs exchange outputs with other GPUs that require them for the next round of computation. While these methods are effective for large-scale graphs, they are not directly compatible with hypergraphs.

\section{Proposed Methods}
\label{sec:methods}
In this section, we present our proposed algorithms for learning from relational databases. We begin by constructing a hypergraph representation of the given relational database, which preserves fine-grained relationships and facilitates hypergraph neural network learning (Section~\ref{subsec:hypergraph_generation}). After transforming the relational database into a hypergraph, we apply our proposed model, rel-HNN, which captures complex relationships at multiple levels of granularity (Section~\ref{subsec:hypergraph_neural_network}). Finally, in Section~\ref{subsec:split_hnn}, we introduce our split-parallel hypergraph learning algorithm designed to scale effectively to large relational databases.

\begin{figure}
    \centering
    \includegraphics[width=0.95\linewidth]{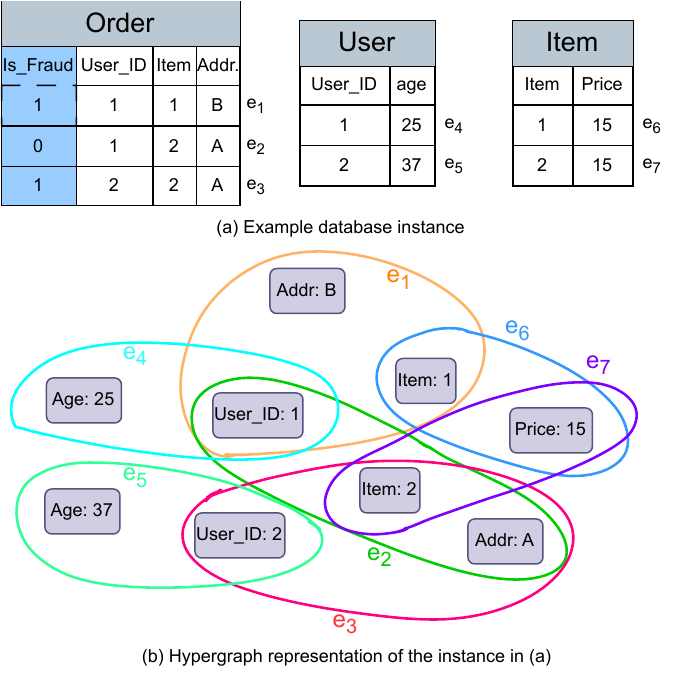}
    \caption{An example hypergraph generation for a relational database}
    \label{fig:hypergraph}
\end{figure}

\begin{algorithm}[]
    \SetAlgoLined
    \SetKwInOut{Input}{Input}
    \SetKwInOut{Output}{Output}
    \Input{$ RDB$: A relational database}
    \Output{$H = (V, E, X)$: A hypergraph}
    \Begin{
    $V, E, \texttt{node\_map} \gets \emptyset, \emptyset, \{\}$ \;

\ForEach{table $T^k \in RDB$}{
  \ForEach{row $i$ in $T^k$}{
    \ForEach{column $j$ in $T^k$}{
      \If{$(\text{Attr}^k_j, T^k_{i,j})$$ \notin$ \texttt{node\_map}}{
        $v \gets new\_node()$ \;
        \texttt{node\_map}[$(\text{Attr}^k_j, T^k_{i,j})$] $\gets v$ \;
        $V \gets V \cup \{v\}$ \;
      }
    }
  }
}

\ForEach{table $T^k \in \mathcal{R}$}{
  \ForEach{row $i$ in $T^k$}{
    $e \gets \emptyset$ \;
    \ForEach{column $j$ in $T^k$}{
      $v \gets$ \texttt{node\_map}[$(\text{Attr}^k_j, T^k_{i,j})$] \;
      $e \gets e \cup \{v\}$ \;
    }
    $E \gets E \cup \{e\}$ \;
  }
}

\ForEach{$v \in V$}{
    $X[v] \gets$ Feature vector of node $v \in V$ \;
}

\Return{$H = (V, E, X)$}
    }

 \caption{Hypergraph Generation}
 \label{algo:algo1}
\end{algorithm}

\subsection{Relational Database to Hypergraph}
\label{subsec:hypergraph_generation}

We now detail our hypergraph construction approach for relational databases, which enables hypergraph neural network learning by capturing complex, fine-grained intra-tuple relationships beyond conventional primary key–foreign key constraints. Our approach decomposes each tuple in the RDB into attribute-value pairs. Instead of representing a tuple or row as a node as existing GNN-based approaches, we create a node for each unique attribute-value pair, $(\text{Attr}^k_j, T^k_{i,j})$, found in all the tables contained by the database (Algorithm \ref{algo:algo1}, Lines 3-9). Then, for each table $T^k \in \text{RDB}$, for each row $T^k_{i,:}$, we create a hyperedge that connects the nodes associated with the attribute-value pairs, $(\text{Attr}^k_j, T^k_{i,j})$, contained by the row (Algorithm \ref{algo:algo1}, Lines 10-16). This hypergraph formulation provides a natural way to model fine-grained interactions within tuples, moving beyond rigid primary key–foreign key constraints, enabling the learning of richer representations. Figure~\ref{fig:hypergraph} illustrates a hypergraph representation for a relational database, where attribute–value pairs are represented as nodes. Edges are shown as colored circles, each connecting more than two nodes to form a hyperedge.

For the feature vector $X_v$ of a node $v \in V$, we consider two techniques (Algorithm~\ref{algo:algo1}, Lines 17–18). In the first approach, we assign one-hot encoded vectors as feature representations for the nodes. In the second approach, we construct the feature vector such that each index corresponds to either an attribute from the tables or a value present in the tables. For each node, we initialize a zero vector and set the indices corresponding to its associated attribute and value to 1.

\begin{figure*}
    \centering
    \includegraphics[width=0.75 \linewidth]{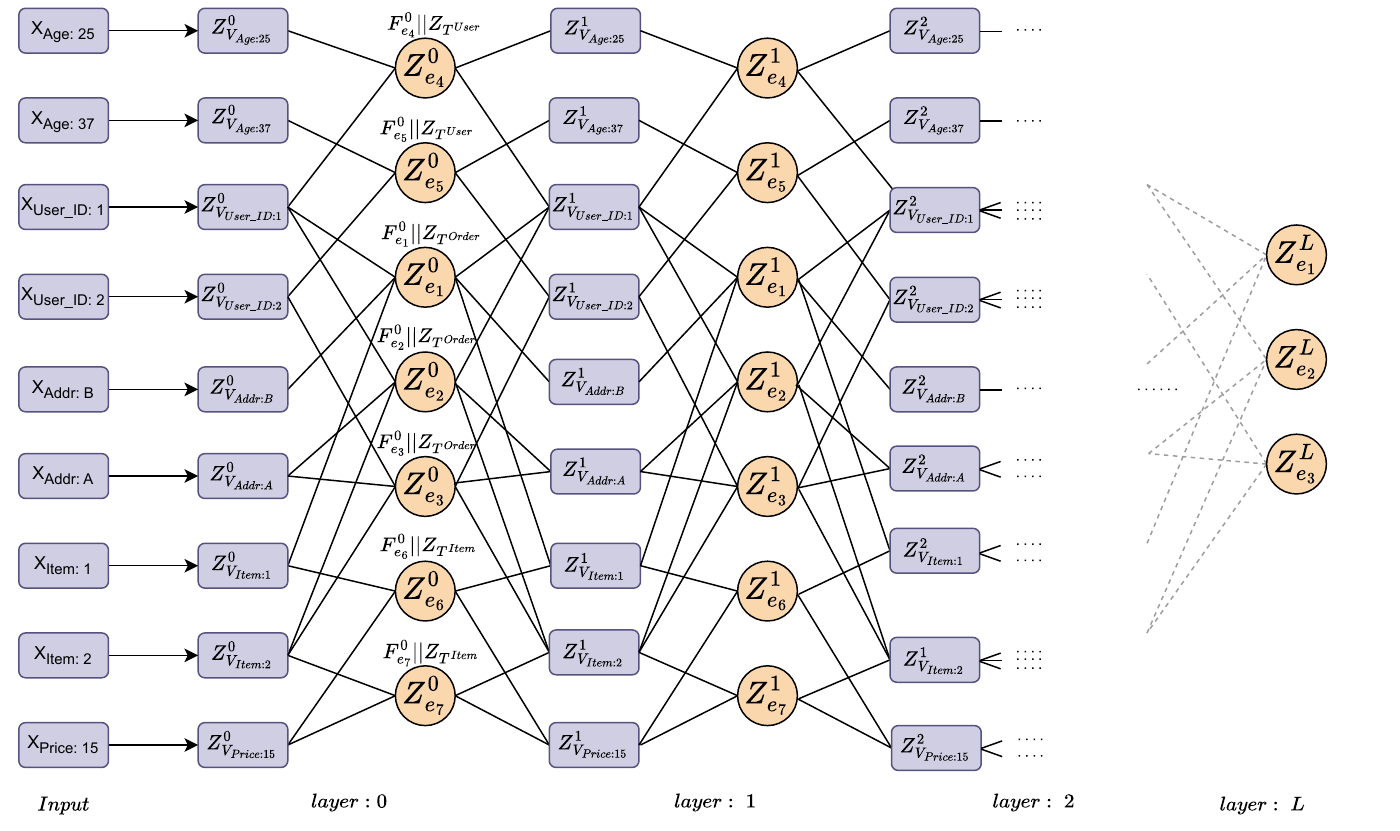}
    \Description{HNN architecture for the hypergraph shown in Figure \ref{fig:hypergraph} (b)}
    \caption{HNN architecture for the hypergraph shown in Figure \ref{fig:hypergraph}}
    \label{fig:HNN}
\end{figure*}

\subsection{rel-HNN: Hypergraph Neural Network for Relational data}
\label{subsec:hypergraph_neural_network}
 
Building on the hypergraph construction described in the previous subsection, we now present rel-HNN, our proposed hypergraph learning algorithm designed to operate on the resulting hypergraph derived from a relational database. Rel-HNN adopts a two-phase message-passing mechanism that alternates between aggregating information from nodes to hyperedges and then from hyperedges back to nodes. In each step of hyperedge embedding, messages (i.e., embeddings) are aggregated from the nodes it connects. Similarly, each node updates its representation by receiving messages from the hyperedges to which it belongs. Figure~\ref{fig:HNN} illustrates the architecture of rel-HNN as applied to the hypergraph depicted in Figure~\ref{fig:hypergraph}.

At the initial layer (layer 0), we compute the embedding $Z^0_v$ of a node $v \in V$ by applying a multilayer perceptron $MLP_V^0$ to its input feature vector $X_v$, as shown in Equation~\ref{eqn:1}. Learning node embeddings that represent attribute–value pairs allows the model to capture fine-grained semantic relationships within tuples, resulting in more expressive representations of relational data.

\begin{equation}
Z^0_v = MLP_V^0(X_v)
\label{eqn:1}
\end{equation}

Subsequently, for each hyperedge $e \in E$, we determine its intermediate embedding at the initial layer, denoted as $F^0_e$, by applying another multilayer perceptron to the sum of the initial node embeddings it connects, as defined in Equation~\ref{eqn:2}.

\begin{equation}
F^0_e = MLP_E^0\left( \sum_{v \in e} Z^0_v \right)
\label{eqn:2}
\end{equation}

In Figure~\ref{fig:HNN}, the initial embedding $F^{0}{e_4}$ of edge $e_4$ is determined by applying the multilayer perceptron on the sum of the initial embeddings of node $V_{Age:25}$ and node $V_{User_ID: 1}$, i.e., $F^{0}{e_4}$ = $MLP^0{E} (Z_{V_{Age:25}}$ + $Z_{V_{User_ID: 1}})$, as edge $e_4$ is connected to these two nodes. By aggregating node embeddings to learn hyperedge embeddings that represent tuples, the model encodes higher-order interactions and co-occurrence patterns among attribute–value pairs, providing a comprehensive understanding of tuple-level semantics. 

For each hyperedge $e \in E$, we learn its final embedding at $layer: 0$ by concatenating its intermediate embedding $F^0_e$ with $Z_{T_e}$, the embedding of the table that contains the row corresponding to the hyperedge (Equation \ref{eqn:3}).

\begin{equation}
Z^0_e = \text{CONCAT}(F^0_e, Z_{T_e})
\label{eqn:3}
\end{equation}
In Figure~\ref{fig:HNN}, the final embedding of edge $e_4$ at layer 0, $Z^{0}{e4}$, is obtained by concatenating the initial embedding of $e_4$, $F^{0}{e4}$, with the table embedding of table $T^{User}$, as $e_4$ originates from the $User$ table; i.e., $Z^{0}{e4} = F^{0}{e4} || Z{T^{User}}$. In our model, learning table-level embeddings enables the incorporation of global patterns shared across all tuples within a table.

\begin{algorithm}[t]
    \SetAlgoLined
    \SetKwInOut{Input}{Input}
    \SetKwInOut{Output}{Output}
    \Input{$H = (V, E, X)$: A hypergraph, 
    $RDB$: A relational database,
    $epochs$: Number of epochs }
    \Output{$\bigcup_{l=0}^{L} \{MLP_{V}^l \cup MLP_{E}^l\}$: The $MLP$ parameters for nodes and hyperedges,
    $\bigcup_{T \in RDB} Z_T$: The table embeddings
    
    }
    \Begin{
        Initialize parameters $\bigcup_{l=0}^{L} \{MLP_{V}^l \cup MLP_{E}^l\}$ and $\bigcup_{T \in RDB} Z_T$ \;
        \For{$\text{epoch} \gets 1$ \KwTo $\text{epochs}$}{
            \For{$v \in V$}{
                $Z^0_v \gets MLP_V^0(X_v)$\;
            
            }
            \For{$e \in E$}{
                $F^0_e \gets MLP_E^0\left( \sum_{v \in e} Z^0_v \right)$\;
                $Z^0_e \gets \text{CONCAT}(F^0_e, Z_{T_e})$\;
            
            }
            \For{$l \gets 1 \to L$}{
                \For{$v \in V$}{
                    $Z^l_v \gets MLP_V^l\left( \sum_{e \in \mathcal{E}_v} Z^{l-1}_e \right)$\;
                
                }
                \For{$e \in E$}{
                $Z^l_e \gets MLP_E^l\left( \sum_{v \in e} Z^l_v \right)$\;
                }
                
            }
            $\mathcal{L} \gets loss\_function(\cup_{e\in E}Z_e^L, RDB)$\;
            Update the parameters of $\bigcup_{l=1}^{L} MLP_{V}^l$ , $\bigcup_{l=1}^{L} MLP_{E}^l$ and $\bigcup_{T \in RDB} Z_T$ to minimize $\mathcal{L}$;
        }
    }
 \caption{rel-HNN}
 \label{algo:algo2}
\end{algorithm}

\begin{figure*}[t]
    \centering
    \includegraphics[width=\linewidth]{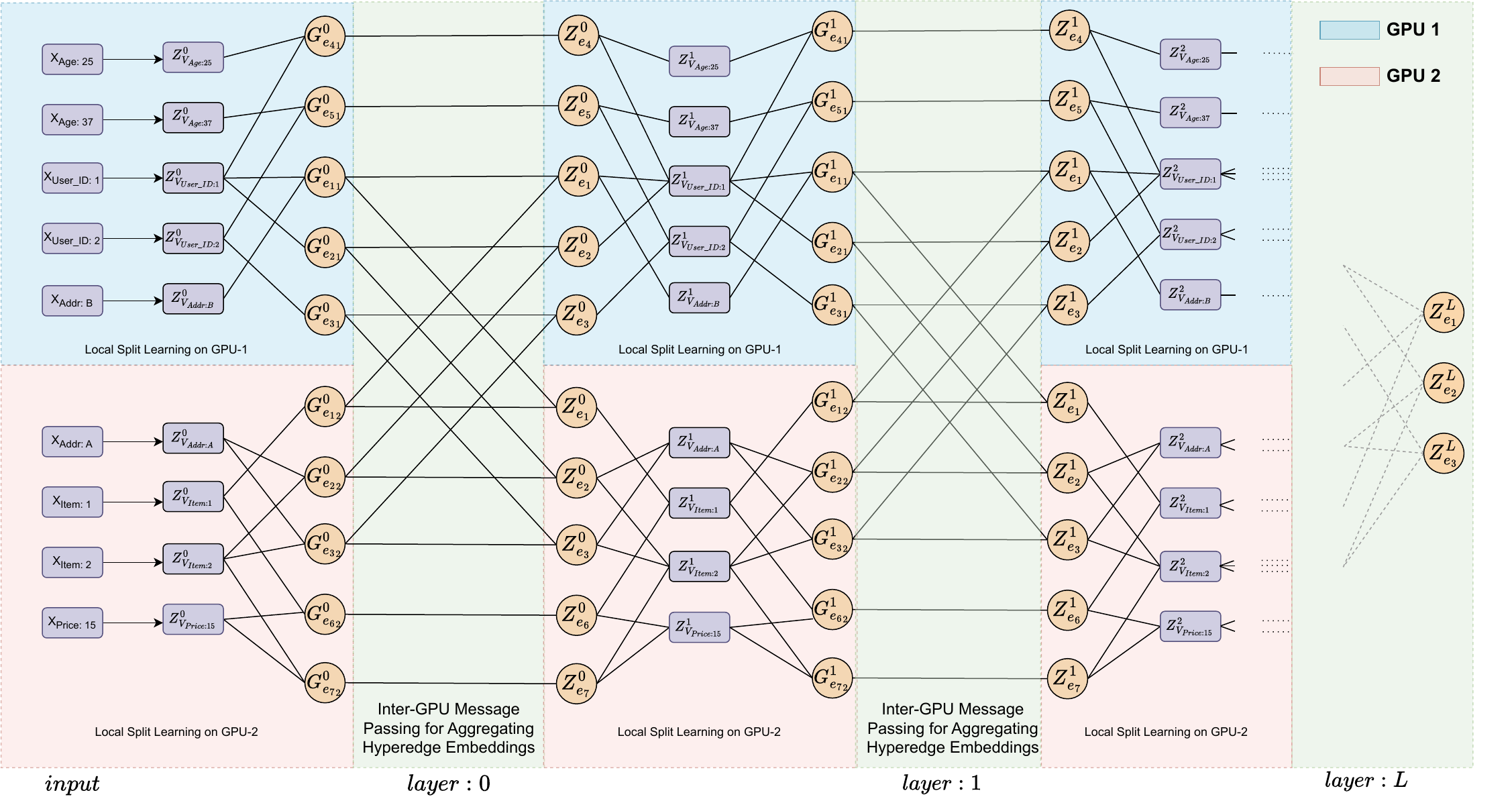}
    \caption{Split-Parallel learning of Hypergraph Neural Network (HNN) on two different GPUs. Blue and red shaded area represent the local processes of GPU-1 and GPU-2 respectively.}
    \label{fig:split_learning}
\end{figure*}

For the intermediate layers (layer 1 to layer L), we determine the embedding of a node $v$, $Z^l_v$, by aggregating the hyperedge embeddings from the previous layer (Equation \ref{eqn:4}). Here, $MLP_v^l$ is the multi-layer perceptron for nodes at layer $l$ and $\mathcal{E}_v$ is the set of hyperedges in $E$ that contains $v$.

\begin{equation}
Z^l_v = MLP_V^l\left( \sum_{e \in \mathcal{E}_v} Z^{l-1}_e \right)
\label{eqn:4}
\end{equation}

For node $V_{User\_{id}: 1}$ in Figure~\ref{fig:HNN}, the embedding at layer 1 is determined by applying the multilayer perceptron $MLP^{0}_V$ to the summation of the final embeddings at layer 0 of hyperedges $e_4$, $e_1$, and $e_2$, as these hyperedges include node $V_{User\_{id}: 1}$.

Similarly, we determine the embedding of a hyperedge $e$ at layer $l$, denoted as $Z^l_e$, by aggregating the node embeddings from the previous layer, as defined in Equation~\ref{eqn:5}. Here, $MLP_E^l$ refers to the multilayer perceptron applied to hyperedges at layer $l$. Note that, through shared attribute–value pairs, the model is able to capture complex relationships between tuples both within and across tables. 

\begin{equation}
Z^l_e = MLP_E^l\left( \sum_{v \in e} Z^l_v \right)
\label{eqn:5}
\end{equation}

For each hyperedge $e$ corresponding to a row in the target table $T^{tg}$, the final embedding $Z^L_e$, where $L$ is the last layer, represents the predicted class probability. In Figure~\ref{fig:HNN}, the hyperedges $e_1$, $e_2$, and $e_3$ correspond to rows of the target table $T^{Order}$.

Algorithm \ref{algo:algo2} presents the pseudocode for the rel-HNN algorithm. The algorithm takes the hypergraph $H = (V, E, X)$, the relational database $RDB$, and the epoch number as input and provides model parameters after training as output. The algorithm first initializes the learnable parameters $\bigcup_{l=0}^{L} \{MLP_{V}^l \cup MLP_{E}^l\}$ both for the nodes and the hyperedges, respectively. In addition, for each table $T \in \text{RDB}$, we initialize an embedding, $Z_T$, which are learnable parameters of the model (Algorithm \ref{algo:algo2}, Line 2). After determining the class probabilities for the hyperedges corresponding to the tuples in the target table as discussed (Algorithm \ref{algo:algo2}, Lines 4-13), the parameters are updated using backpropagation based on the loss determined previously (Algorithm \ref{algo:algo2}, Line 15).

\subsection{Split-Parallel Hypergraph Learning for large databases}
\label{subsec:split_hnn}

To enable scalable learning on large hypergraphs derived from real-world relational databases, we propose a split-parallel hypergraph neural network algorithm. In practice, relational databases tend to be large, and converting them into hypergraphs results in a substantial number of nodes and hyperedges. In particular, assigning a node to each attribute–value pair significantly increases the number of nodes, especially in sparse datasets. Learning on such large hypergraphs introduces critical challenges in terms of computational runtime and memory usage, often making it infeasible to load the entire hypergraph into the limited memory of a single GPU. To address these limitations, we partition the hypergraph by dividing the node set into disjoint splits and distribute the computation of each split and its incident hyperedges across multiple GPUs. This strategy effectively mitigates both memory and runtime bottlenecks, making the training process practical for large-scale relational databases. However, learning from node-partitioned hypergraphs is non-trivial, as hyperedge embeddings rely on aggregating node embeddings that may be distributed across multiple GPUs, leading to communication overheads and synchronization challenges during training.

To utilize parallel processing, assuming there are $N$ GPUs available, we divide the nodes $V$ in the hypergraph $H = (V, E, X)$ into $N$ partitions, $V_1, V_2,$ ... , and $ V_N$. Following the partition, we split the hypergraph $H = (V, E, X)$ into $N$ partitions, $H_1 = (V_1, E_1, X_1)$, $H_2 = (V_2, E_2, X_2)$, ... , and $H_N = (V_N, E_N, X_N)$. Here, $E_i = \{\{v | v\in e \text{ and } v \in V_i \} | e \in E\text{ and } e \cap V_i \neq \emptyset \}$. Given $N$ GPUs ($GPU_1, GPU_2,$, ..., and $GPU_N$), we load the hypergraph $H_i$ into $GPU_i$. Now, in each GPU, $GPU_i$, in parallel, we determine the embedding of a node $v\in V_i$ at layer 0, $Z_v^0$, by applying the same multilayer perceptron layer $MLP_V^0$ to the feature vector $X_v$, similar to Equation \ref{eqn:1}. In Figure \ref{fig:split_learning}, we demonstrate an example of parallel learning using two GPUs (GPU-1 and GPU-2). We divide the nodes in two partition, node $V_1 = \{V_{Age:25}, V_{Age:37}, V_{User\_ID:1},V_{User|ID:2}, V_{Addr: B}\}$ and $V_2 = \{V_{Addr: A}, V_{Item: 1}, V_{Item: 2}, V_{Price: 15}\}$. The local computations are visualized in the blue-shaded area for $GPU_1$ and the red-shaded area for $GPU_2$. For each node, the embedding at layer 0 is learned locally on the assigned GPU. For example, the embedding for $V_{Age:25}$ is learned in GPU-1. 

\begin{algorithm}[]
    \SetAlgoLined
    \SetKwInOut{Input}{Input}
    \SetKwInOut{Output}{Output}
    \Input{$H_i = (V_i, E_i, X_i)$: A hypergraph, 
    $RDB$: A relational database,
    $epochs$: Number of epochs }
    \Output{$\bigcup_{l=0}^{L} \{MLP_{V}^l \cup MLP_{E_{\mathcal{L}}}^l \cup MLP_{E_{\sigma}}^l\}$: The $MLP$ parameters for nodes and hyperedges,
    $\bigcup_{T \in RDB} Z_T$: The table embeddings
    
    }
    \Begin{
        Initialize parameters $\bigcup_{l=0}^{L} \{MLP_{V}^l \cup MLP_{E_{\mathcal{L}}}^l \cup MLP_{E_{\sigma}}^l\} $ and $\bigcup_{T \in RDB} Z_T$ \;
        \For{$\text{epoch} \gets 1$ \KwTo $\text{epochs}$}{
            \For{$v \in V_i$}{
                $Z^0_v \gets MLP_V^0(X_v)$\;
            
            }
            \For{$e \in E_i$}{
                $G^0_{e_i} \gets MLP^0_{E_{\mathcal{L}}}\left( \sum_{v \in e} Z^0_v \right)$\;
                Send $G^0_{e_i}$ to $\cup_{j=1, j \neq i}^N GPU_j$\;
                }
            Wait for $\cup_{e \in E_i} \cup_{j=1, j \neq i}^N G^0_{e_j}$\ to arrive \;
            \For{$e \in E_i$}{
                $F^0_e \gets
                MLP^0_{E_{\sigma}}\left( \sum_{i = 0}^N  G^0_{e_i}\right)$\;
                $Z^0_e \gets \text{CONCAT}(F^0_e, Z_{T_e})$\;
            
            }
            \For{$l \gets 1 \to L$}{
                \For{$v \in V_i$}{
                    $Z^l_v \gets MLP_V^l\left( \sum_{e \in \mathcal{E}_v} Z^{l-1}_e \right)$\;
                
                }
                \For{$e \in E_i$}{
                $G^l_{e_i} \gets MLP^l_{E_{\mathcal{L}}}\left( \sum_{v \in e} Z^l_v \right)$ \;
                Send $G^l_{e_i}$ to $\cup_{j=1, j \neq i}^N GPU_j$ \;
                }
                Wait for $\cup_{e \in E_i} \cup_{j=1, j \neq i}^N G^l_{e_j}$\ to arrive \;
                \For{$e \in E_i$}{
                $Z^l_e \gets MLP^l_{E_{\sigma}}\left( \sum_{i = 0}^N  G^l_{e_i}\right)$\;
                }
                
            }
            $\mathcal{L} \gets loss\_function(\cup_{e\in E}Z_e^L, RDB)$\;
            Update the parameters of $\bigcup_{l=1}^{L} MLP_{V}^l$ , $\bigcup_{l=0}^{L}\{ MLP_{E_{\mathcal{L}}}^l \cup MLP_{E_{\sigma}}^l\} $, and $\bigcup_{T \in RDB} Z_T$ to minimize $\mathcal{L}$;
        }
    }
 \caption{Split-Parallel rel-HNN}
 \label{algo:algo3}
\end{algorithm}
In rel-HNN, we apply the multilayer perceptron layer, $MLP_E^0$, to the sum of node embeddings to determine the intermediate embedding at the initial layer, $F^0_e$ (Equation ~\ref{eqn:2}). For split learning, We divide $MLP_E^0$ into a single-hidden-layer MLP, consisting of a linear transformation $MLP^0_{E_{\mathcal{L}}}$ followed by a non-linear activation function $MLP^0_{E_{\sigma}}$. Next, we compute the local hyperedge embeddings at the initial layer associated with the $i$-th partition, denoted as $G^0_{e_i}$, by applying $MLP^0_{E_{\mathcal{L}}}$ to the sum of the node embeddings at layer 0 (Equation ~\ref{eqn:6}).

\begin{equation}
G^0_{e_i} = MLP^0_{E_{\mathcal{L}}}\left( \sum_{v \in e} Z^0_v \right)
\label{eqn:7}
\end{equation}

In Figure \ref{fig:split_learning}, the local hyperedge embedding at layer 0 for hyperedge $e_1$ at GPU-1, $G^0_{{e_1}_1}$, is learned from $V_{User\_ID:1}$ and $V_{Addr: B}$, the constituent nodes in the partition. On the other hand, for GPU-2, $G^0_{{e_1}_2}$ is learned from $V_{Item:1}$.

To learn the global hyperedge embeddings at the initial layer, $F_e^0$, we determine the sum of the local embeddings and apply the non-linear activation function $MLP^0_{E_{\sigma}}$ (Equation ~\ref{eqn:7}). This aggregation of local embeddings from different GPUs requires inter-GPU message passing.

\begin{equation}
F^0_e = MLP^0_{E_{\sigma}}\left( \sum_{i = 0}^N  G^0_{e_i}\right)
\label{eqn:8}
\end{equation}

Finally, for each hyperedge $e \in E$, we learn its final embedding at layer 0, $Z_e^0$, by concatenating $F^0_e$ with $Z_{T_e}$, the embedding of the corresponding table (Equation \ref{eqn:3}). In Figure \ref{fig:split_learning}, the final embedding of $e_1$ at layer 0, $Z_{e_1}^0$, is learned by aggregating their local embeddings at GPU-1 and GPU-2 and then concatenating their corresponding table embedding, $Z_{T^{Order}}$.

For the intermediate layers, we determine the embedding of a node $v \in V_i$, $Z^l_v$, using Equation \ref{eqn:4}, in parallel. The process of determining node embeddings requires no inter-GPU messaging as the corresponding hyperedge embeddings are already accumulated in the assigned GPU. For example, in Figure \ref{fig:split_learning}, the embedding of node $V_{User\_ID:1}$ at layer 1, $Z^1_{V_{User\_ID:1}}$, is calculated by aggregating $Z^0_{e_1}$ and $Z^0_{e_2}$ which are already accumulated in GPU-1. For the hyperedge embeddings, similar to the initial layer, we divide $MLP_E^l$ as a single hidden layer MLP consisting of one linear transformation, $MLP^l_{E_{\mathcal{L}}}$, followed by a non-linear activation function $MLP^l_{\sigma}$. We determine the local hyperedge embeddings at the initial layer associated with the $i$-th partition, $G^l_{e_i}$, by applying $MLP^l_{E_{\mathcal{L}}}$ on the sum of the embeddings of the nodes from the same layer (Equation ~\ref{eqn:9}).
\begin{equation}
G^l_{e_i} = MLP^l_{E_{\mathcal{L}}}\left( \sum_{v \in e} Z^l_v \right)
\label{eqn:9}
\end{equation}

Finally, we learn the global hyperedge embeddings at layer $l$, $Z_e^l$, we determine the sum of the local embeddings and apply the non-linear activation function $MLP^l_{E_{\sigma}}$ (Equation ~\ref{eqn:10}). 
\begin{equation}
Z^l_e = MLP^l_{E_{\sigma}}\left( \sum_{i = 0}^N  G^l_{e_i}\right)
\label{eqn:10}
\end{equation}

\section{Empirical Evaluation}
\label{sec:results}
\begin{table*}[ht]
\centering
\caption{Statistics of Classification Dataset}
\begin{tabular}{lccccccccc}
\hline
\textbf{Statistic} & \textbf{Hepa} & \textbf{Bupa} & \textbf{Pima} & \textbf{Cora} & \textbf{SameGen} & \textbf{st\_loan} & \textbf{Mutag} & \textbf{rel-f1 (top3)} & \textbf{rel-f1 (dnf)} \\
\hline
Number of Nodes & 6488 & 495 & 1773 & 7927 & 141 & 1031 & 13492 & 85263 & 85607 \\
\hline
Number of Hyperedges & 12927 & 2762 & 6912 & 57353 & 1536 & 5288 & 10324 & 76730 & 86742 \\
\hline
Total Number of Tables & 7 & 9 & 9 & 3 & 4 & 10 & 3 & 10 & 10 \\
\hline
Total Number of Rows & 12927 & 2762 & 6912 & 57353 & 1536 & 5288 & 10324 & 76730 & 86742 \\
\hline
Total Number of Columns & 26 & 16 & 18 & 6 & 8 & 15 & 14 & 70 & 70 \\
\hline
Number of Classes & 2 & 2 & 2 & 7 & 2 & 2 & 2 & 2 & 2 \\
\hline
\end{tabular}
\label{tab:dataset_stats}
\end{table*}

\begin{table*}[htbp]
\centering
\caption{Performance comparison of different methods across datasets. Each method's mean AUROC is followed by the standard deviation in parentheses on the next line. Values shown in bold indicate the best performance per dataset.}
\begin{tabular}{lccccccccc}
\hline
\textbf{Method} & \textbf{Hepa} & \textbf{Bupa} & \textbf{Pima} & \textbf{Cora} & \textbf{SameGen} & \textbf{st\_loan} & \textbf{Mutag} & \textbf{rel-f1 (top3)} & \textbf{rel-f1 (dnf)} \\
\hline

\multirow{2}{*}{GCN} 
& 0.5484 & 0.5121 & 0.5438 & 0.5365 & 0.5245 & 0.5872& 0.6357 & 0.5582 & 0.5122 \\
& (0.0820) & (0.0193) & (0.0746) & (0.0502) & (0.0918) & (0.0374) & (0.0650) & (0.0192) & (0.0837) \\
\hline

\multirow{2}{*}{GAT} 
& 0.5451 & 0.5115 & 0.5442 & 0.5393 & 0.5119 & 0.5821 & 0.6320 & 0.5589 & 0.5064 \\
& (0.0465) & (0.0746) & (0.0746) & (0.0529) & (0.0509) & (0.0501) & (0.0683) & (0.0283) & (0.0928) \\
\hline

\multirow{2}{*}{SPARE-GCN} 
& 0.5636 & 0.5128 & 0.5511 & 0.5712 & 0.5233 & 0.5883 & 0.6044 & 0.5657 & 0.5241 \\
& (0.0374) & (0.0837) & (0.0374) & (0.0650) & (0.0465) & (0.0650) & (0.0912) & (0.0928) & (0.0918) \\
\hline

\multirow{2}{*}{SPARE-GAT} 
& 0.5594 & 0.5263 & 0.5527 & 0.5679 & 0.5276 &0.5948 & 0.6271 & 0.5323 & 0.5222 \\
& (0.0465) & (0.0452) & (0.0509) & (0.0370) & (0.0746) & (0.0918) & (0.0928) & (0.0501) & (0.0703) \\
\hline

\multirow{2}{*}{ATJ-net} 
& 0.5950 & 0.4749 & 0.6072 & 0.6242 & 0.5030 & \textbf{0.9411} & 0.8812 & 0.6412 & 0.5030 \\
& (0.0444) & (0.0452) & (0.0314) & (0.0111) & (0.0078) & (0.0097) & (0.0505) & (0.0081) & (0.0078) \\
\hline
\multirow{2}{*}{rel-HNN-one} 
& 0.8264 & 0.5802 & \textbf{0.7158} & \textbf{0.7364} & 0.8096 & 0.6903 & 0.8777 & 0.8602 & 0.7614 \\
& (0.0776) & (0.1086) & (0.0180) & (0.0582) & (0.0642) & (0.0812) & (0.0496) & (0.0193) & (0.0182) \\
\hline
\multirow{2}{*}{rel-HNN-av} 
& 0.6515 & 0.5759 & 0.7097 & 0.7041 & 0.8166 & 0.6443 & \textbf{0.8976} & 0.8553 & 0.7563 \\
& (0.1485) & (0.0896) & (0.0205) & (0.0336) & (0.0641) & (0.1086) & (0.0560) & (0.0118) & (0.0143) \\
\hline
\multirow{2}{*}{rel-HNN-one-t} 
& \textbf{0.8916} & \textbf{0.6233} & 0.7023 & 0.6521 & \textbf{0.8250} & 0.7909 & 0.8697 & \textbf{0.8685} & \textbf{0.7616} \\
& (0.0596) & (0.0928) & (0.0111) & (0.0825) & (0.0646) & (0.0446) & (0.0537) & (0.0263) & (0.0232) \\
\hline
\multirow{2}{*}{rel-HNN-av-t} 
& 0.8667 & 0.6086 & 0.6982 & 0.5119 & 0.8218 & 0.8642 & 0.7300 & 0.7541 & 0.7206 \\
& (0.0287) & (0.0806) & (0.0260) & (0.0239) & (0.0647) & (0.0392) & (0.2495) & (0.1261) & (0.0488) \\
\hline
\end{tabular}
\label{tab:method_comparison_classification_multirow}
\end{table*}

In this section, we present a comprehensive set of experiments focused on finding the effectiveness and performance gain of our proposed rel-HNN approach. The structure of this section is as follows: we begin by detailing the datasets used and addressing model architecture, along with the experimental settings. Next, we separately demonstrate the performance improvements achieved by our approach on both classification and regression tasks. Lastly, we analyze the effectiveness of our split parallel hypergraph learning approach on both relational and hypergraph datasets.

\subsection{Experiment Design}
To verify the effectiveness of our proposed rel-HNN model, we conducted experiments on both classification and regression tasks. We compared its performance against state-of-the-art graph-based algorithms for learning on relational data. Specifically, we applied GCN~\cite{gnn3} and GAT~\cite{gnn4} on graphs constructed from tuples connected through primary key–foreign key (PK–FK) relationships. For SPARE~\cite{paper2}, we considered both GCN- and GAT-based variants, referred to as SPARE-GCN and SPARE-GAT, respectively. We also evaluated ATJ-Net~\cite{paper3}, which leverages a heterogeneous GNN architecture. In our experiments, we included four versions of rel-HNN. Rel-HNN-one uses one-hot encoding, while rel-HNN-av uses $attribute-value$ encoding for node features. In both rel-HNN-one and rel-HNN-av, table embeddings are omitted. By contrast, rel-HNN-one-t and rel-HNN-av-t include learnable table embeddings in their respective architectures.  For all the rel-HNN variants, we have set the number of layers $L = 2$. The embedding length of all the nodes and hyperedges is fixed at two. For rel-HNN-one-t and rel-HNN-av-t, the table embedding dimension is set to 8 across all the datasets. We have adopted stratified 5-fold cross-validation to preserve class distribution. All experiments were conducted on a workstation equipped with an Intel Core i7-7700 CPU @ 3.60GHz, 48GB RAM, and four NVIDIA GeForce RTX 3060 Ti GPUs with 8GB of memory each.

\subsection{Performance on Classification Tasks}
\label{section:4.3}
For the classification task, nine different datasets are selected from various domains: Hepatitis B disease (Hepa), liver disorder (Bupa),  Diabetes disease (Pima), citation networks (Cora), Kinship information (SameGen), Student loan information (st\_loan), Mutagenicity information (Mutag), Formula 1 dataset (rel-f1(top3) \& rel-f1(dnf)). The first 7 datasets are from the CTU Relational Learning Repository \cite{motl2024ctupraguerelationallearning} and the last 2 datasets are from The Relational Deep Learning Benchmark Repository\cite{relbench}. These databases are widely used for evaluating supervised learning models on classification tasks involving relational data, where a categorical attribute is to be predicted. The datasets vary in complexity, ranging from 3 to 10 relational tables and containing up to 70 columns. All datasets are designed for binary classification tasks, except for the Cora dataset, which contains 7 distinct classes. The statistics for the datasets used in classification experiments are presented in Table~\ref{tab:dataset_stats}.

Table~\ref{tab:method_comparison_classification_multirow} presents the AUROC scores for different methods evaluated across the datasets. The results clearly show that the rel-HNN variants consistently outperform the baseline methods with the highest AUROC score on eight out of nine datasets. Rel-HNN variants also maintain lower or comparable standard deviations to other methods, indicating stable performance across folds. Among the state-of-the-art methods, ATJ-net performs the best on six out of the nine datasets. Specifically on the
$st\_loan$ dataset, ATJ-net achieves the highest AUROC score of 0.9411 across all methods. However, on datasets such as $Bupa$, $SameGen$, and \textit{rel-f1(dnf)}, ATJ-net exhibits comparatively lower performance than the GCN, GAT, and SPARE variants. Among the graph-based approaches, SPARE variants generally achieved a slightly higher AUROC score compared to GCN and GAT.  

Among the rel-HNN variants, rel-HNN-one-t (rel-HNN with one-hot encoding and table embedding) achieves the highest AUROC scores on most datasets, including \textit{Hepa} (0.8916), \textit{Bupa} (0.6233), \textit{SameGen} (0.8250), \textit{rel-f1 (top3)} (0.8685), and \textit{rel-f1 (dnf)} (0.7616). Rel-HNN-one-t demonstrates competitive AUROC performance on the remaining datasets as well. The standard deviations for this variant are also relatively low or comparable across datasets, indicating stable and consistent performance. Among the other three versions of rel-HNN, rel-HNN-av-t is a strong contender that outperforms the other two variants on datasets such as \textit{Hepa}, \textit{Bupa}, \textit{SameGen}, and \textit{st\_loan}. The benefit of learning table embeddings explicitly is evident from the performance gains of the \textit{-t} variants. Both \textit{rel-HNN-one} and \textit{rel-HNN-av} exhibit notable improvements after the concatenation of the table embedding that represents the global information of all tuples in a table. An important observation is that in the datasets where the \textit{non-t} variants perform relatively better such as, the number of tables is relatively small, suggesting that explicit table-level embeddings may be less beneficial when the relational structure is simple or shallow. For example, there are only three tables in both \textit{Cora} and \textit{Mutag}, where the \textit{non-t} variants perform better. For the dataset \textit{Pima}, the number of tables is relatively higher (nine), but the performance gap in favor of non-t variants is also insignificant compared to \textit{Cora} and \textit{Mutag}.
However, compared to existing state-of-the-art methods, rel-HNN-one and rel-HNN-av demonstrate substantial performance gains, despite not utilizing table-level embeddings.

The performance results presented in Table~\ref{tab:method_comparison_classification_multirow}, when analyzed alongside the dataset statistics in Table~\ref{tab:dataset_stats}, reveal several key trends. Across a wide range of datasets, including both small-scale ones like \textit{Bupa}, \textit{Pima}, and \textit{SameGen}, and large-scale datasets such as \textit{Hepa}, \textit{Cora}, \textit{Mutag}, \textit{rel-f1 (top3)} and \textit{rel-f1 (dnf)}, our proposed methods demonstrate superior performance. The relative performance gain for our models is relatively higher for the larger datasets such as \textit{Hepa} (49.86\%), \textit{Cora} (17.98\%), \textit{rel-f1 (top3)} (35.45\%) and \textit{rel-f1 (dnf)} (45.32\%) with \textit{Mutag}(1.86\%) being an exception. On the contrary, the relative improvement is moderate for smaller datasets like \textit{Bupa} (18.44\%) and \textit{Pima} (17.88\%), while \textit{SameGen} shows a notably higher gain (56.37\%). Interestingly, although ATJ-net performs well on the \textit{st\_loan} dataset, achieving the highest score of 0.9411, it under performs elsewhere, suggesting limited generalization.

In summary, our methods adapt effectively to varying levels of schema complexity and data volume, making them suitable for both compact and large-scale relational databases. Rel-HNN-one-t emerges as the most robust model, delivering both high AUROC scores and low variance. These results collectively demonstrate the effectiveness of our proposed framework on classification problem across diverse relational datasets.

\subsection{Performance on Regression Tasks} 
\begin{table}[ht]
\centering
\caption{Statistics of Regression Datasets}
\begin{tabular}{lcccc}
\hline
\textbf{Statistic} 
& \shortstack{\textbf{Pyr}} 
& \shortstack{\textbf{CM}} 
& \shortstack{\textbf{Pubs}} 
& \shortstack{\textbf{Bio}} \\
\hline
Number of Nodes           & 130   & 5215  & 681   & 15435 \\
Number of Hyperedges      & 296   & 3864  & 245   & 21895 \\
Total Number of Tables          & 2     & 8     & 10    & 5 \\
Total Number of Rows            & 296   & 3864  & 245   & 21895 \\
Total Number of Columns        & 13    & 59    & 61    & 14 \\
Number of Classes         & 63    & 273   & 12    & 100 \\
\hline
\end{tabular}%
\label{tab:regression_dataset_stats}
\end{table}

\begin{table*}[ht]
\centering
\caption{RMSE Comparison Across Datasets in Regression Tasks}
\begin{tabular}{lcccc}
\toprule
\textbf{Method} & \textbf{Pyrimidine} & \textbf{ClassicModels} & \textbf{Pubs} & \textbf{Biodegradability} \\
\midrule
 
GCN & 0.1195 $\pm$ 0.0591 & 955.4509 $\pm$ 90.8273 & 225.6450 $\pm$ 20.0912 & 18.8374 $\pm$ 5.3645 \\
GAT & 0.1183 $\pm$ 0.0530 & 989.2736 $\pm$ 80.5091 & 231.1827 $\pm$ 16.7465 & 19.5109 $\pm$ 5.1827 \\
SPARE-GCN & 0.1061 $\pm$ 0.0517 & 809.0918 $\pm$ 67.7364 & 125.6509 $\pm$ 12.9283 & 14.3746 $\pm$ 7.6509 \\
SPARE-GAT & 0.1149 $\pm$ 0.0469 & 808.6547 $\pm$ 59.9182 & 142.8374 $\pm$ 15.6501 & 18.0928 $\pm$ 4.9821 \\
ATJ-net & 0.1235 $\pm$ 0.0331 & 728.8391 $\pm$ 52.2696 & 95.3952 $\pm$ 7.1283 & 24.7605 $\pm$ 1.1405 \\
rel-HNN-one & \textbf{0.0792 $\pm$ 0.0394} & 120.7200 $\pm$ 18.9351 & 6.0572 $\pm$ 1.0807 & 1.5909 $\pm$ 0.2871 \\
rel-HNN-av & 0.0843 $\pm$ 0.0308 & 117.6724 $\pm$ 17.4639 & 6.0106 $\pm$ 1.1637 & 1.6729 $\pm$ 0.0198 \\
rel-HNN-one-t & 0.0975 $\pm$ 0.0365 & 117.5294 $\pm$ 9.2348 & \textbf{5.4055 $\pm$ 1.4554} & \textbf{1.4779 $\pm$ 0.4093} \\
rel-HNN-av-t & 0.0979 $\pm$ 0.0574 & \textbf{115.8039 $\pm$ 14.1657} & 5.6332 $\pm$ 1.5954 & 1.6390 $\pm$ 0.3934 \\
\bottomrule
\end{tabular}
\label{tab:rmse_results}
\end{table*}

For regression tasks, we have collected four relational datasets from The CTU Relational Learning Repository\cite{motl2024ctupraguerelationallearning}. Table \ref{tab:regression_dataset_stats} presents the statistical summary of four relational datasets used for regression tasks.  The datasets exhibit significant variation in size and complexity. For instance, \textit{Biodegradability (Bio)} is the largest in terms of both nodes (15,435) and hyperedges (21,895), suggesting a rich relational structure and potentially complex learning dynamics. In contrast, \textit{Pyrimidine (Pyr)} is the smallest, with only 130 nodes and 296 hyperedges, making it a lightweight dataset suitable for rapid experimentation or benchmarking. The \textit{ClassicModels (CM)} dataset shows a high number of tables (8), columns (59), and classes (273), indicating a detailed schema and a fine-grained prediction task. Similarly, the \textit{Pubs} dataset, despite having a relatively small number of rows (245), has a large number of columns (61), which pose challenges related to feature sparsity or redundancy.

Table~\ref{tab:rmse_results} reports the Root Mean Square Error (RMSE) performance across four regression datasets for various methods. Across all datasets, the proposed rel-HNN variants significantly outperform the state-of-the-art methods in terms of RMSE values. On the \textit{Pyrimidine} dataset, which is relatively small, rel-HNN-one achieves the lowest RMSE of $0.0792$, outperforming SPARE-GCN, the best-performing baseline method, by a substantial margin. Other variants of rel-HNN have also performed better than the existing approaches. In the \textit{ClassicModels} dataset, which contains more tables and a larger schema, the ATJ-net performs the best among existing methods with an RMSE value of $728.8391$. In contrast, the rel-HNN variants reduce the error by significantly, with rel-HNN-av-t achieving an RMSE value of $115.8039$. This demonstrates the ability of our models to generalize better in complex relational structures. The improvements are even more pronounced in the \textit{Pubs} dataset. Here, ATJ-net yields an RMSE of $95.3952$, while all rel-HNN versions reduce the error drastically to around $6$, with rel-HNN-one-t achieving the lowest RMSE of $5.4055$. This suggests that the rel-HNN architecture is particularly effective in datasets with rich attribute columns and intricate relational dependencies. For the largest dataset, \textit{Biodegradability}, \textit{SPARE-GCN} achieves the best performance among prior approaches, with an RMSE of $24.7605$. In contrast, \textit{rel-HNN-one-t} significantly outperforms all baselines, achieving the lowest RMSE of $1.4779$. Other \textit{rel-HNN} variants also attain RMSE values around $1.5$, demonstrating the scalability and effectiveness of our method in handling large and complex relational datasets. Similar to classification problem, we can observe that the relative performance gain is higher for larger and complex datasets. For larger and complex datasets such as \textit{ClassicModels} (97.83\%), \textit{Pubs} (94.34\%), and \textit{Biodegradability} (94.03\%), the performance gain exceeds 90\%, while for the smaller \textit{Pyrimidine} dataset, it is comparatively lower at 86.63\%.

Overall, the rel-HNN models consistently achieve lower RMSE values and demonstrate greater stability across datasets, as indicated by smaller standard deviations in most cases. This underlines both the accuracy and reliability of the proposed approach for regression tasks over heterogeneous relational data.
The different variants of rel-HNN demonstrate unique strengths. Rel-HNN-one achieves the lowest RMSE score on \textit{Pyrimidine}, which consists of two tables only, suggesting its effectiveness for limited schema complexity. On the other hand, for larger and complex datasets with a higher number of tables, the table embedding variants (one-t and av-t) tend to perform better. Rel-HNN-one-t provides the best performance on \textit{Pubs} (5.4055) and \textit{Biodegradability} (1.4779) and near-best performance on \textit{ClassicModels} (117.5294). Rel-HNN-av-t achieves the best performance on \textit{ClassicModels} (115.8039) and is narrowly outperformed on \textit{Pubs} (5.6332) by \textit{rel-HNN-one-t}. For \textit{ClassicModels} dataset, although rel-HNN-av-t is outperformed by both rel-HNN-one and rel-HNN-one-t, it performs better than its non-table counterpart, rel-HNN-av. Overall, the flexibility among variants allows the rel-HNN framework to adapt effectively across a broad spectrum of relational learning tasks.

\subsection{Split-Parallel Hypergraph Learning Performance}


To observe the effectiveness of parallel processing, we conducted experiments using the split learning process on a classification task. All experiments were performed using the rel-HNN-one-t version of the rel-HNN model. We evaluated performance on two different sets of datasets. The first set consists of the same datasets used in Section~\ref{section:4.3} (Table~\ref{tab:dataset_stats}). The second set includes widely used benchmark hypergraph datasets. 

\begin{figure}
    \centering
\begin{minipage}[b]{0.32\linewidth}
\begin{tikzpicture}
\begin{axis}[
    title={Hepa},
    height=3cm,
    ybar,
    bar width=8pt,
    enlarge y limits=false,
    enlarge x limits=0.15,
    ylabel={Time (ms)},
    y label style={at={(axis description cs:-0.30,0.0)},anchor=west},
    xlabel={$N$},
    x label style={at={(axis description cs:-0.3,-0.1)},anchor=north west},
    symbolic x coords={1, 2, 3, 4},
    xtick=data,
    ymin=17,
    ymax=22
]
\addplot+[fill=blue!80] coordinates {
    (1, 21.41)
    (2, 20.37)
    (3, 19.96)
    (4, 18.98)
};
\end{axis}
\end{tikzpicture}
\end{minipage}
\begin{minipage}[b]{0.32\linewidth}
\begin{tikzpicture}
\begin{axis}[
    title={Bupa},
    height=3cm,
    ybar,
    bar width=8pt,
    enlarge y limits=false,
    enlarge x limits=0.15,
    ylabel={Time (ms)},
    y label style={at={(axis description cs:-0.30,0.0)},anchor=west},
    xlabel={$N$},
    x label style={at={(axis description cs:-0.3,-0.1)},anchor=north west},
    symbolic x coords={1, 2, 3, 4},
    xtick=data,
    ymin=13,
    ymax=18
]
\addplot+[fill=blue!80] coordinates {
    (1, 17.43)
    (2, 15.78)
    (3, 15.65)
    (4, 14.34)
};
\end{axis}
\end{tikzpicture}
\end{minipage}
\begin{minipage}[b]{0.32\linewidth}
\begin{tikzpicture}
\begin{axis}[
    title={Pima},
    height=3cm,
    ybar,
    bar width=8pt,
    enlarge y limits=false,
    enlarge x limits=0.15,
    ylabel={Time (ms)},
    y label style={at={(axis description cs:-0.30,0.0)},anchor=west},
    xlabel={$N$},
    x label style={at={(axis description cs:-0.3,-0.1)},anchor=north west},
    symbolic x coords={1, 2, 3, 4},
    xtick=data,
    ymin=5,
    ymax=16
]
\addplot+[fill=blue!80] coordinates {
    (1, 9.44)
(2, 14.48)
(3, 15.68)
(4, 13.96)
};
\end{axis}
\end{tikzpicture}
\end{minipage}

\begin{minipage}[b]{0.32\linewidth}
\begin{tikzpicture}
\begin{axis}[
    title={Cora},
    height=3cm,
    ybar,
    bar width=8pt,
    enlarge y limits=false,
    enlarge x limits=0.15,
    ylabel={Time (ms)},
    y label style={at={(axis description cs:-0.30,0.0)},anchor=west},
    xlabel={$N$},
    x label style={at={(axis description cs:-0.3,-0.1)},anchor=north west},
    symbolic x coords={1, 2, 3, 4},
    xtick=data,
    ymin=15,
    ymax=60
]
\addplot+[fill=blue!80] coordinates {
(1, 58.25)
(2, 33.57)
(3, 25.92)
(4, 21.18)
};
\end{axis}
\end{tikzpicture}
\end{minipage}
\begin{minipage}[b]{0.32\linewidth}
\begin{tikzpicture}
\begin{axis}[
    title={SameGen},
    height=3cm,
    ybar,
    bar width=8pt,
    enlarge y limits=false,
    enlarge x limits=0.15,
    ylabel={Time (ms)},
    y label style={at={(axis description cs:-0.30,0.0)},anchor=west},
    xlabel={$N$},
    x label style={at={(axis description cs:-0.3,-0.1)},anchor=north west},
    symbolic x coords={1, 2, 3, 4},
    xtick=data,
    ymin=5,
    ymax=16
]
\addplot+[fill=blue!80] coordinates {
(1, 9.84)
(2, 15.29)
(3, 14.25)
(4, 14.02)
};
\end{axis}
\end{tikzpicture}
\end{minipage}
\begin{minipage}[b]{0.32\linewidth}
\begin{tikzpicture}
\begin{axis}[
    title={st\_loan},
    height=3cm,
    ybar,
    bar width=8pt,
    enlarge y limits=false,
    enlarge x limits=0.15,
    ylabel={Time (ms)},
    y label style={at={(axis description cs:-0.30,0.0)},anchor=west},
    xlabel={$N$},
    x label style={at={(axis description cs:-0.3,-0.1)},anchor=north west},
    symbolic x coords={1, 2, 3, 4},
    xtick=data,
    ymin=5,
    ymax=16
]
\addplot+[fill=blue!80] coordinates {
(1, 12.29)
(2, 15.65)
(3, 14.33)
(4, 14.22)
};
\end{axis}
\end{tikzpicture}
\end{minipage}

\begin{minipage}[b]{0.32\linewidth}
\begin{tikzpicture}
\begin{axis}[
    title={Mutag},
    height=3cm,
    ybar,
    bar width=8pt,
    enlarge y limits=false,
    enlarge x limits=0.15,
    ylabel={Time (ms)},
    y label style={at={(axis description cs:-0.30,0.0)},anchor=west},
    xlabel={$N$},
    x label style={at={(axis description cs:-0.3,-0.1)},anchor=north west},
    symbolic x coords={1, 2, 3, 4},
    xtick=data,
    ymin=12,
    ymax=25
]
\addplot+[fill=blue!80] coordinates {
(1, 23.60)
(2, 19.17)
(3, 16.07)
(4, 16.73)
};
\end{axis}
\end{tikzpicture}
\end{minipage}
\begin{minipage}[b]{0.32\linewidth}
\begin{tikzpicture}
\begin{axis}[
    title={rel-f1 (top3)},
    height=3cm,
    ybar,
    bar width=8pt,
    enlarge y limits=false,
    enlarge x limits=0.15,
    yticklabel style={xshift=2pt,font=\small},
    ylabel={Time (ms)},
    y label style={at={(axis description cs:-0.3,0.0)},anchor=west},
    xlabel={$N$},
    x label style={at={(axis description cs:-0.3,-0.1)},anchor=north west},
    symbolic x coords={1, 2, 3, 4},
    xtick=data,
    ymin=100,
    ymax=700
]
\addplot+[fill=blue!80] coordinates {
(1, 645.83)
(2, 309.80)
(3, 244.67)
(4, 208.63)
};
\end{axis}
\end{tikzpicture}
\end{minipage}
\begin{minipage}[b]{0.32\linewidth}
\begin{tikzpicture}
\begin{axis}[
    title={rel-f1 (dnf)},
    height=3cm,
    ybar,
    bar width=8pt,
    enlarge y limits=false,
    enlarge x limits=0.15,
    yticklabel style={xshift=2pt,font=\small},
    ylabel={Time (ms)},
    y label style={at={(axis description cs:-0.30,0.0)},anchor=west},
    xlabel={$N$},
    x label style={at={(axis description cs:-0.3,-0.1)},anchor=north west},
    symbolic x coords={1, 2, 3, 4},
    xtick=data,
    ymin=100,
    ymax=700
]
\addplot+[fill=blue!80] coordinates {
(1, 672.33)
(2, 298.14)
(3, 245.28)
(4, 211.42)
};
\end{axis}
\end{tikzpicture}
\end{minipage}
    \caption{Training time per epoch (in milliseconds) across relational databases for different numbers of GPUs ($N$)}
    \label{fig:split_learning1}
\end{figure}
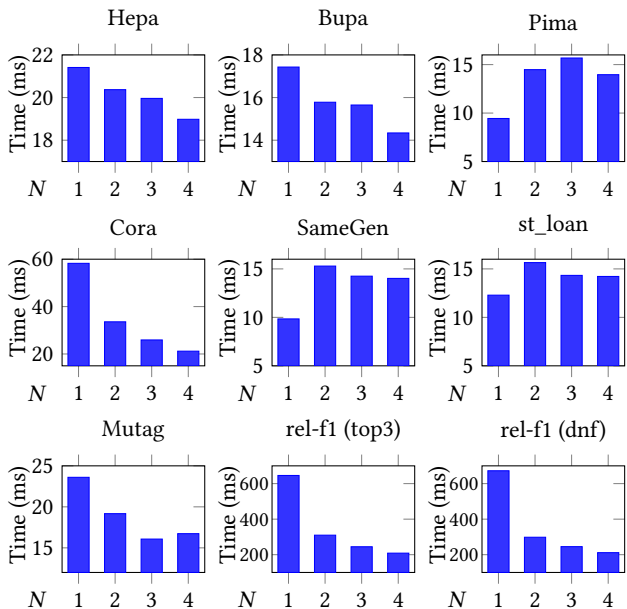
Figure \ref{fig:split_learning1} presents the training time per epoch (in milliseconds) across multiple datasets under varying numbers of GPUs ($N$), reflecting the impact of split-learning and parallelism. A clear trend is observed for larger datasets as \textit{Hepa}, \textit{Cora}, \textit{Mutag}, \textit{rel-f1 (top3)} and \textit{rel-f1 (dnf)}. As we increase the number of GPUs, the training time per epoch decreases substantially. For example, on the largest \textit{rel-f1(dnf)} dataset, the training time per epoch drops significantly from 672.33 ms with a single GPU to 211.42 ms with four GPUs—a speedup of nearly 3.18x. This indicates the substantial benefit of parallel training for large datasets. The speedup tends to diminish as the dataset size decreases, primarily due to insufficient workload to fully utilize multiple GPUs, leading to communication overheads diminishing the benefits of parallelization. For example, on dataset \textit{Mutag}, which is relatively smaller than \textit{rel-f1(dnf)}, the speedup drops to 1.41x, with the training time per epoch decreasing from 23.60 ms to 16.73 ms. On much smaller datasets such as \textit{Pima}, \textit{SameGen}, and \textit{st\_loan}, we observe that increasing the number of GPUs increases the training times per epoch as we employ two GPUs. As the number of GPUs increases, the training time per epoch decreases, but fails to improve beyond the single-GPU performance. Here, the inter-GPU communication and processing overheads for aggregating hyperedge embeddings are outweighing the benefits of parallelization. However, despite being a smaller dataset, on \textit{Bupa}, our algorithm achieves a modest speedup of 1.22x.


In our experiments, we have analyzed the performance of our split-parallel algorithm on three benchmark hypergraph datasets that are widely used for hypergraph-based tasks. Table~\ref{Table: Hypergraph Statistics} presents the structural statistics of the datasets: \textit{Citeseer}, \textit{DBLP}, and \textit{Pubmed}. Among them, \textit{DBLP} is the largest in terms of both the number of nodes (41,302) and hyperedges (22,363), indicating a highly complex and densely connected hypergraph structure. \textit{Pubmed} represents a medium-sized dataset with 19,717 nodes and 7,963 hyperedges, whereas \textit{Citeseer} is the smallest, comprising 3,312 nodes and 1,079 hyperedges. Despite its relatively small scale, \textit{Citeseer} has the highest feature dimensionality, with each node represented by a 3,703-dimensional feature vector. In contrast, \textit{Pubmed} and \textit{DBLP} exhibit more moderate feature lengths of 500 and 1,425, respectively. The diversity across these datasets in terms of size and feature complexity makes them well-suited for evaluating the effectiveness of the proposed split-parallel hypergraph learning approach.

\begin{table}[htbp]
\centering
\caption{Statistics of Benchmark Hypergraph Datasets}
\begin{tabular}{lccc}
\toprule
\textbf{Property} & \textbf{DBLP} & \textbf{Pubmed} & \textbf{Citeseer} \\
\midrule
Number of Nodes & 41,302 & 19,717 & 3,312 \\
Number of Hyperedges & 22,363 & 7,963 & 1,079 \\
Length of Feature Vector & 1,425 & 500 & 3,703 \\
\bottomrule
\end{tabular}
\label{Table: Hypergraph Statistics}
\end{table}
Figure \ref{fig:split_learning_benchmark_hypergraph} presents the per-epoch training time for \textit{Citeseer}, \textit{DBLP}, and \textit{Pubmed} across different values of $N$, representing the number of GPUs used for parallel split-learning. The most significant runtime reduction is observed on the largest dataset, \textit{DBLP}, where training time drops from 98.51 ms with a single GPU to 33.53 ms with four GPUs, achieving a speedup of nearly 2.94$\times$. A similar decreasing trend is seen in \textit{Pubmed}, which shows a consistent improvement as $N$ increases, reducing training time from 20.06 ms to 15.48 ms with a speedup of 1.30$\times$. However, in the case of \textit{Citeseer}, an increase in $N$ results in a rise in training time, likely due to parallelization overhead surpassing the computational benefits for smaller datasets. These findings highlight that while multi-GPU split-learning effectively reduces training time for large-scale datasets, it may introduce diminishing or even negative returns for smaller hypergraphs with limited computational load.

\begin{figure}
    \centering

\begin{minipage}[b]{0.32\linewidth}
\begin{tikzpicture}
\begin{axis}[
    title={DBLP},
    height=3cm,
    ybar,
    bar width=8pt,
    enlarge y limits=false,
    enlarge x limits=0.15,
    yticklabel style={xshift=2pt,font=\small},
    ylabel={Time (ms)},
    y label style={at={(axis description cs:-0.30,0.0)},anchor=west},
    xlabel={$N$},
    x label style={at={(axis description cs:-0.3,-0.1)},anchor=north west},
    symbolic x coords={1, 2, 3, 4},
    xtick=data,
    ymin=10,
    ymax=110
]
\addplot+[fill=blue!80] coordinates {
(1, 98.51)
(2, 56.88)
(3, 42.43)
(4, 33.53)
};
\end{axis}
\end{tikzpicture}
\end{minipage}
\begin{minipage}[b]{0.32\linewidth}
\begin{tikzpicture}
\begin{axis}[
    title={Pubmed },
    height=3cm,
    ybar,
    bar width=8pt,
    enlarge y limits=false,
    enlarge x limits=0.15,
    ylabel={Time (ms)},
    y label style={at={(axis description cs:-0.30,0.0)},anchor=west},
    xlabel={$N$},
    x label style={at={(axis description cs:-0.3,-0.1)},anchor=north west},
    symbolic x coords={1, 2, 3, 4},
    xtick=data,
    ymin=12,
    ymax=21
]
\addplot+[fill=blue!80] coordinates {
(1, 20.06)
(2, 17.75)
(3, 16.31)
(4, 15.48)
};
\end{axis}
\end{tikzpicture}
\end{minipage}
\begin{minipage}[b]{0.32\linewidth}
\begin{tikzpicture}
\begin{axis}[
    title={Citeseer},
    height=3cm,
    ybar,
    bar width=8pt,
    enlarge y limits=false,
    enlarge x limits=0.15,
    ylabel={Time (ms)},
    y label style={at={(axis description cs:-0.30,0.0)},anchor=west},
    xlabel={$N$},
    x label style={at={(axis description cs:-0.3,-0.1)},anchor=north west},
    symbolic x coords={1, 2, 3, 4},
    xtick=data,
    ymin=4,
    ymax=16
]
\addplot+[fill=blue!80] coordinates {
(1, 7.91)
(2, 13.67)
(3, 13.33)
(4, 13.58)
};
\end{axis}
\end{tikzpicture}
\end{minipage}

    \caption{Training time per epoch (in milliseconds)
across hypergraph datasets for different numbers of GPUs ($N$)}
    \label{fig:split_learning_benchmark_hypergraph}
\end{figure}
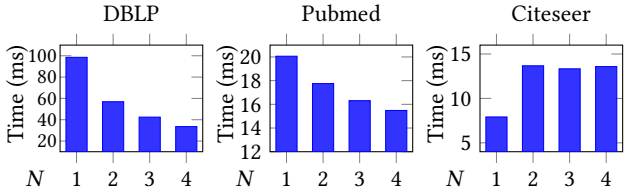

\begin{figure}
    \centering

\begin{minipage}[b]{0.32\linewidth}
\begin{tikzpicture}
\begin{axis}[
    title={|V|=5k, |E|=10k},
    height=3cm,
    ybar,
    bar width=8pt,
    enlarge y limits=false,
    enlarge x limits=0.15,
    yticklabel style={xshift=2pt,font=\small},
    ylabel={Time (ms)},
    y label style={at={(axis description cs:-0.30,0.0)},anchor=west},
    xlabel={$N$},
    x label style={at={(axis description cs:-0.3,-0.1)},anchor=north west},
    symbolic x coords={1, 2, 3, 4},
    xtick=data,
    ymin=8,
    ymax=17
]
\addplot+[fill=blue!80] coordinates {
(1, 10.02)
(2, 13.69)
(3, 13.98) 
(4, 15.28)

};
\end{axis}
\end{tikzpicture}
\end{minipage}
\begin{minipage}[b]{0.32\linewidth}
\begin{tikzpicture}
\begin{axis}[
    title={|V|=5k, |E|=50k},
    height=3cm,
    ybar,
    bar width=8pt,
    enlarge y limits=false,
    enlarge x limits=0.15,
    ylabel={Time (ms)},
    y label style={at={(axis description cs:-0.30,0.0)},anchor=west},
    xlabel={$N$},
    x label style={at={(axis description cs:-0.3,-0.1)},anchor=north west},
    symbolic x coords={1, 2, 3, 4},
    xtick=data,
    ymin=12,
    ymax=32
]
\addplot+[fill=blue!80] coordinates {
(1, 30.52)
(2, 22.40)
(3, 17.26)
(4, 15.37)
};
\end{axis}
\end{tikzpicture}
\end{minipage}
\begin{minipage}[b]{0.32\linewidth}
\begin{tikzpicture}
\begin{axis}[
    title={|V|=5k, |E|=100k},
    height=3cm,
    ybar,
    bar width=8pt,
    enlarge y limits=false,
    enlarge x limits=0.15,
    ylabel={Time (ms)},
    y label style={at={(axis description cs:-0.30,0.0)},anchor=west},
    xlabel={$N$},
    x label style={at={(axis description cs:-0.3,-0.1)},anchor=north west},
    symbolic x coords={1, 2, 3, 4},
    xtick=data,
    ymin=20,
    ymax=60
]
\addplot+[fill=blue!80] coordinates {
(1, 57.25)
(2, 35.39)
(3, 26.34)
(4, 24.55)
};
\end{axis}
\end{tikzpicture}
\end{minipage}
\begin{minipage}[b]{0.32\linewidth}
\begin{tikzpicture}
\begin{axis}[
    title={|V|=10k, |E|=10k},
    height=3cm,
    ybar,
    bar width=8pt,
    enlarge y limits=false,
    enlarge x limits=0.15,
    yticklabel style={xshift=2pt,font=\small},
    ylabel={Time (ms)},
    y label style={at={(axis description cs:-0.30,0.0)},anchor=west},
    xlabel={$N$},
    x label style={at={(axis description cs:-0.3,-0.1)},anchor=north west},
    symbolic x coords={1, 2, 3, 4},
    xtick=data,
    ymin=12,
    ymax=18
]
\addplot+[fill=blue!80] coordinates {
(1, 16.19)
(2, 13.84)
(3, 13.98)
(4, 14.22)
};
\end{axis}
\end{tikzpicture}
\end{minipage}
\begin{minipage}[b]{0.32\linewidth}
\begin{tikzpicture}
\begin{axis}[
    title={|V|=10k, |E|=50k},
    height=3cm,
    ybar,
    bar width=8pt,
    enlarge y limits=false,
    enlarge x limits=0.15,
    ylabel={Time (ms)},
    y label style={at={(axis description cs:-0.30,0.0)},anchor=west},
    xlabel={$N$},
    x label style={at={(axis description cs:-0.3,-0.1)},anchor=north west},
    symbolic x coords={1, 2, 3, 4},
    xtick=data,
    ymin=15,
    ymax=60
]
\addplot+[fill=blue!80] coordinates {
(1, 55.35)
(2, 34.67)
(3, 24.75)
(4, 20.57)
};
\end{axis}
\end{tikzpicture}
\end{minipage}
\begin{minipage}[b]{0.32\linewidth}
\begin{tikzpicture}
\begin{axis}[
    title={|V|=10k, |E|=100k},
    height=3cm,
    ybar,
    bar width=8pt,
    enlarge y limits=false,
    enlarge x limits=0.15,
    yticklabel style={xshift=2pt,font=\small},
    ylabel={Time (ms)},
    y label style={at={(axis description cs:-0.30,0.0)},anchor=west},
    xlabel={$N$},
    x label style={at={(axis description cs:-0.3,-0.1)},anchor=north west},
    symbolic x coords={1, 2, 3, 4},
    xtick=data,
    ymin=30,
    ymax=120
]
\addplot+[fill=blue!80] coordinates {
(1, 109)
(2, 58.95)
(3, 44.54)
(4, 36.04)
};
\end{axis}
\end{tikzpicture}
\end{minipage}

    \caption{Training time per epoch (in milliseconds)
across synthetic hypergraphs for different numbers of GPUs ($N$)}
    \label{fig:synth_results}
\end{figure}
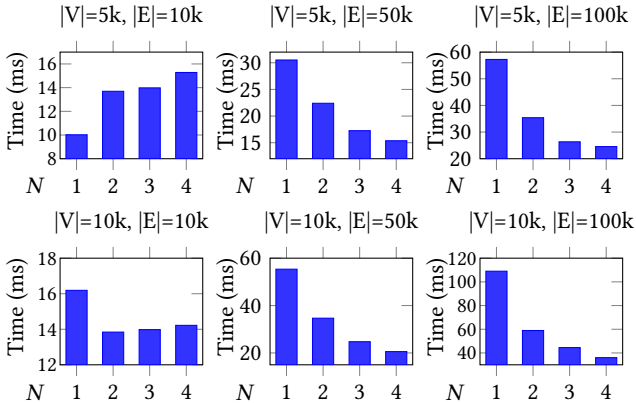

In addition to real-life benchmark hypergraphs, we have also experimented with synthetic hypergraphs generated by varying the number of nodes and hyperedges. For each hyperedge, nodes are drawn from a uniform distribution, where the number of nodes varies from three to ten. The feature vectors and labels are also randomly assigned to each node, where the feature vector length is set to 1024. We varied the number of nodes, $|V|$, between 5,000 and 10,000, and the number of hyperedges, $|E|$, between 10,000 and 100,000 to assess performance across different hypergraph scales. In Figure \ref{fig:synth_results}, we present the per-epoch training time across
the synthetic hypergraphs for different numbers of GPUs. As the size of the hypergraphs increases—either in terms of the number of nodes or hyperedges— the advantages of parallel learning become more pronounced. For the smallest hypergraph with $|V|$=5,000 and $|E|$=10,000, increasing the number of GPUs leads to an increase in training time, due to synchronization overheads. However, as the number of hyperedges is increased to 50,000 and 100,000, the algorithm achieves an increasing speedup of 1.98$\times$ and 2.33$\times$, respectively. For the hypergraph with $|V| = 10{,}000$ and $|E| = 10{,}000$, the training time decreases from 16.19 ms (with a single GPU) to 13.84 ms when using two GPUs. Interestingly, as more GPUs are employed, the training time begins to increase, reaching 14.22 ms with 1.13$\times$ speedup for four GPUs, indicating underutilization of parallel resources. Again, increasing the number of hyperedges to 50,000 and 100,000 results in higher speedups of 2.69$\times$ and 3.02$\times$, respectively. A similar trend is observed with an increasing number of nodes. For instance, with $|E| = 100{,}000$, increasing the number of nodes from 5,000 to 10,000 enhances the speedup from 2.33$\times$ to 3.02$\times$. These results demonstrate that while synchronization overheads can limit speedup gains on smaller hypergraphs, the proposed parallel learning approach achieves substantial performance improvements as the size and complexity of the hypergraphs scale.

\section{Conclusions and Future Work}
In this paper, we presented rel-HNN, a novel hypergraph neural network framework for learning on relational databases. By representing attribute-value pairs as nodes and tuples as hyperedges, our model captures intricate, fine-grained relationships within and across tuples without relying on schema-specific constraints like primary key–foreign key (PK–FK) relationships. Rel-HNN introduces a multi-level embedding strategy to learn representations at the attribute, tuple, and table levels, offering a comprehensive and expressive approach to relational data modeling. To address the scalability challenge posed by large hypergraphs, we further proposed a split-parallel learning algorithm that effectively distributes the workload across multiple GPUs. Our empirical evaluation shows that rel-HNN consistently outperforms state-of-the-art methods in predictive performance and offers significant computational speedups through parallel training. Building on this foundation, federated hypergraph learning and learning on relational data in cloud environments can be promising directions for future research.


\bibliographystyle{ACM-Reference-Format}
\bibliography{bibliography}

\end{document}